\documentclass[twocolumn,prl,aps]{revtex4}
\usepackage{amsmath}   
\usepackage{amssymb}
\usepackage{amsfonts}
\usepackage{graphicx}

\begin{document}
\title
{Magnetoplasma excitations of two-dimensional anisotropic heavy fermions in AlAs quantum wells}

\author{V.~M.~Muravev$^{a}$, A.~R.~Khisameeva$^{a,b}$, V.~N.~Belyanin$^{a,b}$, I.~V.~Kukushkin$^{a}$, L.~Tiemann$^{c}$, C.~Reichl$^{c}$,  W.~Dietsche$^{c}$, W.~Wegscheider$^{c}$}
\affiliation{$^a$ Institute of Solid State Physics RAS, Chernogolovka 142432, Russia\\
$^b$ Moscow Institute of Physics and Technology, Dolgoprudny 141700, Russia\\ 
$^c$ Solid State Physics Laboratory, ETH Zurich, Schafmattstrasse 16, 8093 Zurich, Switzerland \\}
\date{\today}

\date{\today}

\begin{abstract}
The spectra of plasma and magnetoplasma excitations in a two-dimensional system of anisotropic heavy fermions were investigated for the first time. The spectrum of microwave absorption by disk-like samples of stressed AlAs quantum wells at low electron densities showed two plasma resonances separated by a frequency gap. These two plasma resonances correspond to electron mass principle values of $(1.10 \pm 0.05) m_0$ and $(0.20 \pm 0.01) m_0$. The observed results correspond to the case of a single valley strongly anisotropic Fermi surface. It was established that electron density increase results in population of the second valley, manifesting itself as a drastic modification of the plasma spectrum. We directly determined the electron densities in each valley and the inter-valley splitting energy from the ratio of the two plasma frequencies.

\end{abstract}

\maketitle

The last few decades have witnessed a surge in research on the fascinating and often unexpected collective states arising from strong electron-electron interaction. Selectively doped semiconductor heterostructures appeared to be nearly ideal systems for such research owing to their strongly reduced disorder. Examples of phenomena caused by electron correlations include the fractional quantum Hall effect~\cite{Tsui:82}, metal-insulator transition~\cite{Pudalov}, and spin-textured structures~\cite{Spin}. The electron-electron interaction strength is characterized by the ratio of the Coulomb interaction energy to the Fermi energy, and is proportional to the effective mass of charge carriers. This has motivated interest in new two-dimensional electron systems (2DES) that show heavy fermions behavior. One of the most promising materials of choice is n-AlAs~\cite{Shayegan:06}. 

AlAs is an indirect gap semiconductor in which the electrons occupy three equivalent valleys at the $X$ points of the Brillouin zone. This degeneracy is lifted when the electrons are confined to a 2D layer. In a quantum well grown on a GaAs $(001)$ wafer, only the in-plane $[100]$ ($X$) and $[010]$ ($Y$) valleys are occupied for well widths greater than $5$~nm~\cite{Maezawa:92}. This differs from Si  $(001)$ metal oxide semiconductor field-effect transistors (MOSFETs), in which the two valleys with the out-of-plane major axes are occupied. Such unusual behavior stems from biaxial compression of the AlAs layer, induced by lattice mismatch between the AlAs and GaAs. Moreover, the residual in-plane strain lifts the $X$ and $Y$ valley degeneracy, leading to inter-valley energy splitting $\Delta E$ (Fig.~1)~\cite{Ando, Shayegan:02, Shayegan:05, Grayson:11}. This splitting modifies the plasma spectrum, as observed in the present paper. Transport measurements revealed large and anisotropic AlAs electron effective masses $m_{\rm l}=(1.1 \pm 0.1) m_0$ and $m_{\rm tr}=(0.20 \pm 0.02) m_0$~\cite{Smith:87, CR:93, Mass:04}, corresponding to the longitudinal and transverse Fermi ellipsoid axes directions, respectively. The effective Lande $g$-factor of electrons in bulk AlAs ($g^{\ast} = 2$) is much larger than in GaAs ($g^{\ast} = -0.44$). These characteristics make AlAs heterostructure 2DES unique and versatile subject of many-body and valleytronics phenomena investigations.  

Microwave magnetospectroscopy is the most direct method to characterize Fermi surfaces and determine effective masses~\cite{Dresselhaus:54}. This method has revealed well-studied 2D plasma excitations in isotropic single component GaAs heterostructure 2DES~\cite{Stern:67, Allen:77}. All attempts to study plasma dynamics in anisotropic 2DES, however, have been limited to experiments in which an applied magnetic field creates a small anisotropy in the an initially isotropic 2DES~\cite{Barke:86, Kozlov1, Kozlov2}. The only example of collective behavior in a multi-component 2DES was found in GaAs double quantum wells occupied by electrons in one well and holes in the other~\cite{Kukushkin:11}. Heretofore, 2D plasmon physics in systems with native strong mass anisotropy, as well as multi-component 2DES, has not been well explored, despite a number of interesting physical predictions~\cite{Chaplik}. AlAs 2DES provide an important research opportunity by combining strong anisotropy with the ability to tune carrier density in each valley.

Measurements were carried out on high-quality $15$~nm AlAs quantum well heterostructures fabricated by molecular beam epitaxy (MBE) on a $(001)$ GaAs substrate. The electron density $n_s$ and electron mobility $\mu$ were in the ranges of $1{.}7\times 10^{11} - 2{.}4\times 10^{11}\,\text{cm}^{-2}$ and $1{.}2\times 10^{5} - 2{.}0\times 10^{5} \,\text{cm}^{2}/\text{V} \cdot \text{s}$, respectively. A variation of the electron density was achieved by short illumination of the sample. The illumination was performed by a green light emitting diode ($2.2$~eV) at $T = 1.6$~K. A coplanar waveguide (CPW) was fabricated on top of the crystal surface using standard photolithography tools. The waveguide contained a central $1{.}1$~mm wide stripe spaced $0{.}6$~mm from the grounded planes (Fig.~1 inset). The total length of the coplanar waveguide was $4$~mm. The parameters of the waveguide were chosen to provide a characteristic impedance of $Z_0 = 50$~$\Omega$. Six equidistant 2DES disks of diameter $d=0{.}5$~mm were fabricated in the slots of the CPW. The disk centers were spaced $1{.}5$~mm apart to minimize cross-talk effects. Arrows indicate the basic crystallography directions in Fig.~1. We detected the resonant absorption of microwave probe radiation ($f=1-40$~GHz) propagating along the CPW. An alternating electric field concentrated in the slots of the CPW oscillates the 2D plasmas in the disks. Resonant absorption of microwaves occurs whenever a plasmon is excited in a disk. The sample was immersed in a cryostat with a superconducting coil. $50$~$\Omega$ coaxial cables connected the sample in series between a microwave generator ($f=0 - 40$~GHz) and a tunnel diode with a preamplifier placed outside the cryostat. The output power of the generator did not exceed $100$~nW and the output signal was detected by a standard lock-in technique. The magnetic field was applied normal to the surface of the sample. Helium vapor was pumped out to attain a temperature of $T = 1.6$~K.  

Figure~1 shows the magnetic-field dependencies of the coplanar waveguide transmission for several microwave frequencies. The horizontal axis is located at the signal level when no microwave radiation is supplied to the CPW. Each curve shows a resonance with respect to zero magnetic field. Most of the resonances are symmetric, the rest have an asymmetric line-shape. Observed asymmetric resonances in the transmission can be treated and analyzed as Fano-type resonances (see Supplementary Material \textrm{II}~\cite{Fano}). The resonance shifts to lower magnetic fields with increasing microwave frequency $f$, indicating the edge magnetoplasma (EMP) nature of the observed resonance. For frequencies above $15$~GHz, a second plasma resonance arises (inset in Fig.~1). The resonance behavior exhibits the positive magnetodispersion characteristic of a cyclotron magnetoplasmon. For more transmission curves we refer to Fig.~1S in the Supplementary Material \textrm{I}. 

\begin{figure}[!t]
\includegraphics[width=0.47 \textwidth]{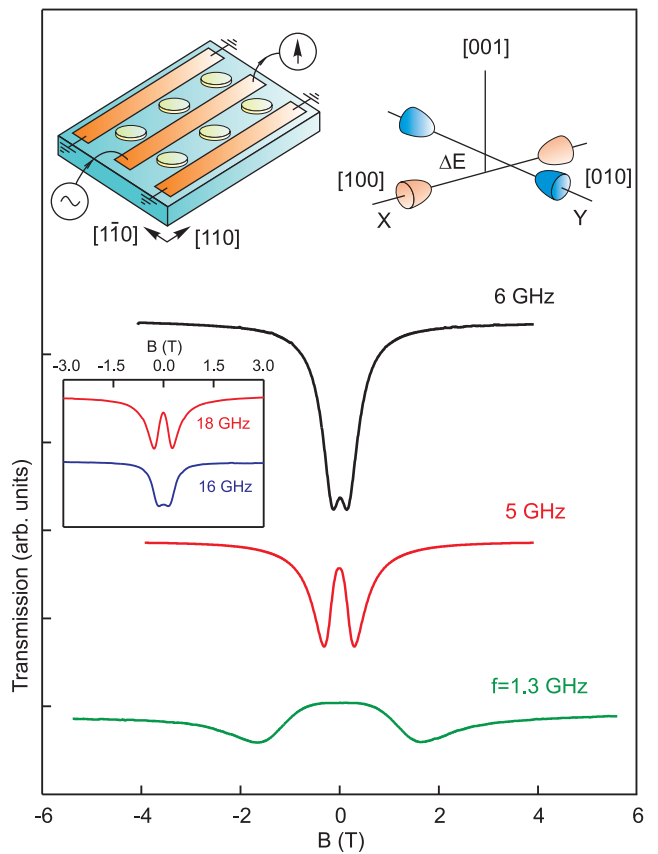}
\caption{Magnetic-field dependencies of the coplanar waveguide transmission at microwave frequencies of $1.3$~GHz, $5$~GHz, and $6$~GHz. Each curve shows a well-defined resonance corresponding to edge magnetoplasmon excitation. Residual in-plane strain in semiconductor structure causes a rise of energy gap $\Delta E$ between the $X$ and $Y$ valleys. The inset shows CPW transmission at $16$~GHz and $18$~GHz. The resonances exhibit positive magnetodispersion inherent to cyclotron magnetoplasmon. The electron density is $n_s = 1{.}7\times 10^{11}\,\text{cm}^{-2}$ at $T=1.5$~K. Schematic drawings of the coplanar waveguide indicating the basic crystallography directions and the AlAs Fermi surface are shown in the upper left and right, respectively.}
\label{1}
\end{figure}     

The resonance origins are best identified by plotting the resonant magnetic-fields versus microwave frequencies, as shown in Fig.~2. The data were obtained at an electron density of $1{.}7\times 10^{11}\,\text{cm}^{-2}$. The magnetodispersion has two branches separated by a frequency gap. The low-frequency branch corresponds to an edge magnetoplasmon propagating along the edge of the disk. This is a mode with anomalously weak attenuation that propagates in a narrow strip near the edge of the 2DES~\cite{Volkov:88, Allen:83}. The EMP frequency decreases as $\omega_{-} \approx \sigma_{xy} q \propto n_s q/B$ in the strong magnetic field limit. The high-frequency branch has a positive magnetodispersion. The electric field $\vec{E}$ aligned along the $[1\bar{1}0]$ (Fig.~1) crystal direction can be factorized into two components along the Fermi ellipsoid axes as $\vec{E}=\vec{E}_{\rm l} + \vec{E}_{\rm tr}$. In the $B=0$~T limit, each of these components excites a separate 2D plasma wave with corresponding masses $m_{\rm l}$ and $m_{\rm tr}$. Therefore, the gap in the magnetoplasmon spectrum of the disk vividly demonstrates the highly anisotropic nature of the Fermi surface in AlAs 2DES~\cite{Dahl, Kozlov2}. For the sake of comparison, we performed the same measurements on a geometrically identical sample made from GaAs quantum well ($d=0.5$~mm, $n_s = 1{.}4\times 10^{11}\,\text{cm}^{-2}$). The inset in Fig.~2 shows that for $B=0$~T the magnetic field edge and cyclotron magnetoplasma modes are degenerate, highlighting the mass $m^{\ast}=0.067 m_0$ isotropism in GaAs.           
\begin{figure}[!t]
\includegraphics[width=0.47 \textwidth]{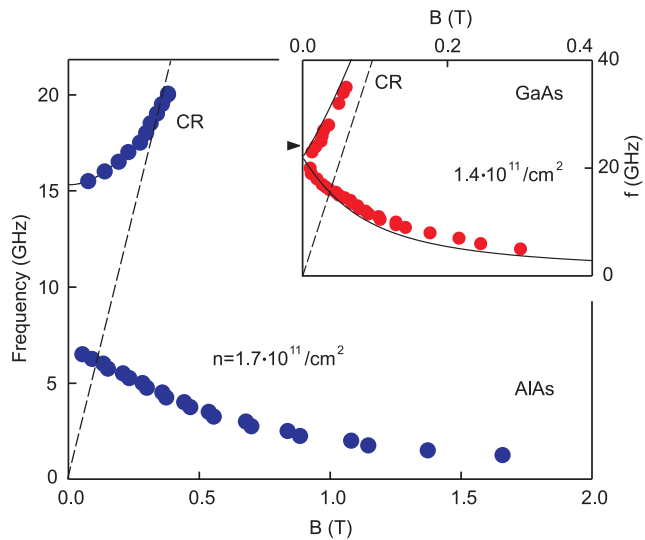}
\caption{Magnetodispersion of two-dimensional plasma excitations  in AlAs disks with anisotropic charge carriers ($n_s=1{.}7\times 10^{11}\,\text{cm}^{-2}$). The plasmon spectrum shows two plasma resonance branches separated by a frequency gap. Inset shows dispersion of magnetoplasmon waves in a GaAs quantum well for electrons with isotropic mass ($n_s=1{.}4\times 10^{11}\,\text{cm}^{-2}$). No frequency gap is observed. The same disk and CPW geometry was used in both cases.}
\label{2}
\end{figure}  

The plasma excitation spectrum in a 2DES with mass anisotropy can be described using the dipole approximation as~\cite{Dahl, Heitmann, Margulis}:
\begin{equation}     
\omega_{\rm l,\rm tr}=\frac{1}{2} \left[ \sqrt{(\Omega_{\rm tr}+\Omega_{\rm l})^2+\omega_c^2} \pm \sqrt{(\Omega_{\rm tr}-\Omega_{\rm l})^2+\omega_c^2} \right],
\label{1}
\end{equation}         
where $\Omega_{\rm l}$ and $\Omega_{\rm tr}$ are plasma frequencies along the main crystallographic directions for $B=0$~T, and $\omega_c=eB/m_c$ is the cyclotron frequency. The cyclotron mass is determined as a geometric mean of effective masses along the crystallographic axes, $m_c=\sqrt{m_{\rm l} m_{\rm tr}}$. The frequencies $\Omega_{\rm l, \rm tr}$ obey the 2D-plasmon dispersion~\cite{Stern:67}:
\begin{equation}     
\Omega_{\rm l,\rm tr}^2=\frac{n_s e^2}{2 m_{\rm l,\rm tr} \varepsilon_0 \varepsilon^{\ast}} q,
\label{2}
\end{equation}      
where $\varepsilon^{\ast}=(\varepsilon_{\rm GaAs}+1)/2$ is the effective dielectric permittivity of the surrounding medium and $q=2{.}4/d$ is the wave vector for the disk geometry~\cite{Kukushkin:03}. From our experiments, we measure zero-field plasma frequencies $\Omega_{\rm l}=(6.5 \pm 0.2)$~GHz and $\Omega_{\rm tr}=(15.3 \pm 0.5)$~GHz. Using Eq.~(2) we find the effective masses in the AlAs quantum well along the main crystallographic directions to be $m_{\rm l}=(1.10 \pm 0.05) m_0$ and $m_{\rm tr}=(0.20 \pm 0.01) m_0$. These mass values agree with results obtained from commensurability oscillation measurements~\cite{Shayegan:06}.

\begin{figure}[!t]
\includegraphics[width=0.47 \textwidth]{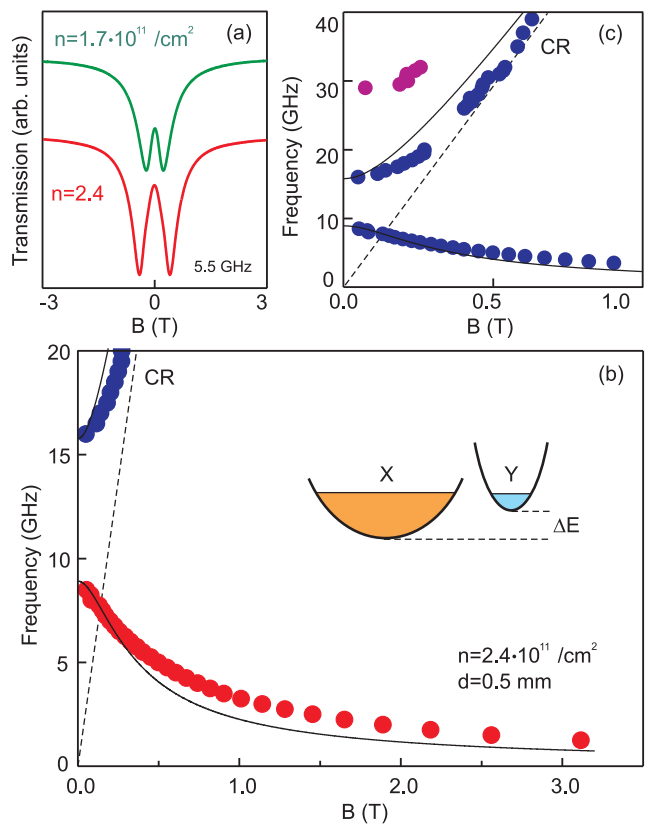}
\caption{(a) Magnetic-field dependencies of the coplanar waveguide transmission for microwave frequency $f=5.5$~GHz with two different 2DES densities. (b) Dispersion of two-dimensional magnetoplasma excitations in AlAs quantum well at $n_s=2{.}4\times 10^{11}\,\text{cm}^{-2}$. Solid line represents theoretical prediction according to Eq.~(1). Schematic drawing of electron spectrum for $n_s=2{.}4\times 10^{11}\,\text{cm}^{-2}$. The $Y$ valley starts to be filled at this density. (c) Extended magnetodispersion of the plasmon modes.}
\label{3}
\end{figure}    

Figure~3(a) shows CPW microwave transmission as a function of magnetic field for the identical samples with 2DES densities of $1{.}7\times 10^{11}\,\text{cm}^{-2}$ and $2{.}4\times 10^{11}\,\text{cm}^{-2}$. A short light flash from a light emitting diode varied the electron density. The magnetoplasma resonance shifted to larger magnetic field values with increased electron density. This is consistent with Eq~($1-2$). However, the zero-field plasma frequencies determined from the detailed magnetodispersion curve of the $2{.}4\times 10^{11}\,\text{cm}^{-2}$ 2DES have a ratio $\Omega_{\rm tr}/\Omega_{\rm l} = (1.80 \pm 0.05)$. This number is inconsistent with Eq.~(\ref{2}), which predicts $\Omega_{\rm tr}/\Omega_{\rm l} = \sqrt{m_{\rm l}/m_{\rm tr}} = (2.3 \pm 0.1)$. This suggests that the plasma dynamics undergo a qualitative metamorphosis when the electron density changes.

We attribute observed phenomenon to the energy splitting between the $X$ and $Y$ valleys. Indeed, the residual in-plane strain lifts the $X$ and $Y$ valley degeneracy, leading to an inter-valley energy splitting $\Delta E$ (Fig.~1). For a 2DES where $n_s=1{.}7\times 10^{11}\,\text{cm}^{-2}$, we find that all electrons occupy only the $X$ valley, leaving the $Y$ valley empty. As we increase the density, some of the electrons begin to fill the $Y$ valley (Fig.~3(b)). The total density is then defined as $n_s=n_x+n_y$, where $n_x$ and $n_y$ are the charge carrier concentrations in the $X$ and $Y$ valleys respectively. The collective plasma excitations in such a system could be considered using a two-component anisotropic plasma model~\cite{Vitlina}. The plasma frequencies along the $[100]$ and $[010]$ directions are described by the following expressions: 

\begin{equation}     
\Omega_{[100]}^2=\frac{e^2 q}{2 \varepsilon_0 \varepsilon^{\ast}} \left( \frac{n_x}{m_{\rm l}}+\frac{n_{y}}{m_{\rm tr}} \right), 
\label{3}
\end{equation} 

\begin{equation}     
\Omega_{[010]}^2=\frac{e^2 q}{2 \varepsilon_0 \varepsilon^{\ast}} \left( \frac{n_x}{m_{\rm tr}}+\frac{n_{y}}{m_{\rm l}} \right). 
\label{4}
\end{equation}

Using these expressions with our obtained plasma frequencies $\Omega_{[100]}$, $\Omega_{[010]}$ and AlAs masses $m_{\rm l}$, $m_{\rm tr}$, we deduced the densities $n_x$ and $n_y$ in each of the valleys to be $n_x=(2{.}10 \pm 0.05)\times 10^{11}\,\text{cm}^{-2}$ and $n_y=(0{.}30 \pm 0.05)\times 10^{11}\,\text{cm}^{-2}$. From the difference of densities $\Delta n = n_x - n_y$, we can directly determine the inter-valley energy splitting $\Delta E$ using the 2D density of states:
\begin{equation}     
\Delta E=\frac{\pi \hbar^2 \Delta n}{\sqrt{m_{\rm l}m_{\rm tr}}}.  
\label{5}
\end{equation}

This calculation gives $\Delta E=(0.90 \pm 0.05)$~meV, which is consistent with all previous studies of valley splitting in AlAs~\cite{Ando, Shayegan:02, Shayegan:05}. However, these studies were conducted in strong magnetic fields ($B > 1$~T), leaving the low magnetic field range unexplored. Our experiments directly determine the valley populations from the plasma frequencies in the weak magnetic field limit. 

The solid lines in Figs.~3(b)~and~(c) represent theoretical predictions of Eq.~(1) with $\Omega_{\rm l} = \Omega_{[100]}$ and $\Omega_{\rm tr} = \Omega_{[010]}$ for the mode magnetodispersion. Some discrepancy exists between the experimental data and theory at moderate magnetic fields. The difference is especially pronounced for the high-frequency cyclotron magnetoplasma mode (Fig.~3(c)). One possible explanation could be hybridization between the cyclotron magnetoplasma mode and an inter-component cyclotron mode. The additional inter-component mode is a dispersionless excitation with frequency $\omega=eB/\sqrt{m_{\rm l} m_{\rm tr}}$, which coincides with the cyclotron resonance~\cite{Chaplik}. Another possible explanation of the observed discrepancy is that $\Omega_{[100]}$, $\Omega_{[010]}$ and the corresponding $\Delta E$ are not constant with the $B$-field. This discrepancy should motivate further research. The agreement between theory and experiment returns in the limit of strong magnetic fields. 

Our results suggest opportunities for future applications of plasmonics in AlAs 2DESs. They can be used to study the inter-valley energy spacing at $B=0$~T using the discovered plasma resonance method. Such experiments could unveil roles of electron-electron interaction in semiconductor valley-splitting~\cite{Ando, Shayegan:02, Shayegan:05}. They could also be used to study novel relativistic plasma excitations~\cite{Muravev}. Relativistic plasma waves in highly anisotropic two-component electron liquids of AlAs may reveal unpredicted physical phenomena. 

The authors gratefully acknowledge financial support from the Russian Scientific Fund (Grant No.~14-12-00693).



\end{document}


\title{Supplementary Material for\\ ``Magnetoplasma excitations of two-dimensional anisotropic heavy fermions in AlAs quantum wells''}
\author{V.~M.~Muravev$^{a}$, A.~R.~Khisameeva$^{a,b}$, V.~N.~Belyanin$^{a,b}$, I.~V.~Kukushkin$^{a}$, L.~Tiemann$^{c}$, C.~Reichl$^{c}$,  W.~Dietsche$^{c}$, W.~Wegscheider$^{c}$}
\affiliation{$^a$ Institute of Solid State Physics RAS, Chernogolovka 142432, Russia\\
$^b$ Moscow Institute of Physics and Technology, Dolgoprudny 141700, Russia\\ 
$^c$ Solid State Physics Laboratory, ETH Zurich, Schafmattstrasse 16, 8093 Zurich, Switzerland \\}

\date{\today}\maketitle

\section{\textrm{I}. ADDITIONAL PICTURES FOR EDGE MAGNETOPLASMA RESONANCE}

In FIG.~1S(a) and FIG.~1S(b) we present the magnetic field dependence of the coplanar microwave transmission at a serious of frequencies $f$. The data of FIG.~1S(a) were obtained at an 2DES electron density of $n_s=1{.}7\times 10^{11}\,\text{cm}^{-2}$ (before illumination), and FIG.~1S(b) for $n_s=2{.}4\times 10^{11}\,\text{cm}^{-2}$ (after illumination). All curves show a resonance that shifts to the lower magnetic field as the frequency is increased. The resonance corresponds to excitation of edge magnetoplasmon (EMP)  propagating along the edge of the disks.   

\begin{figure}[!h]
\includegraphics[width=0.94 \textwidth]{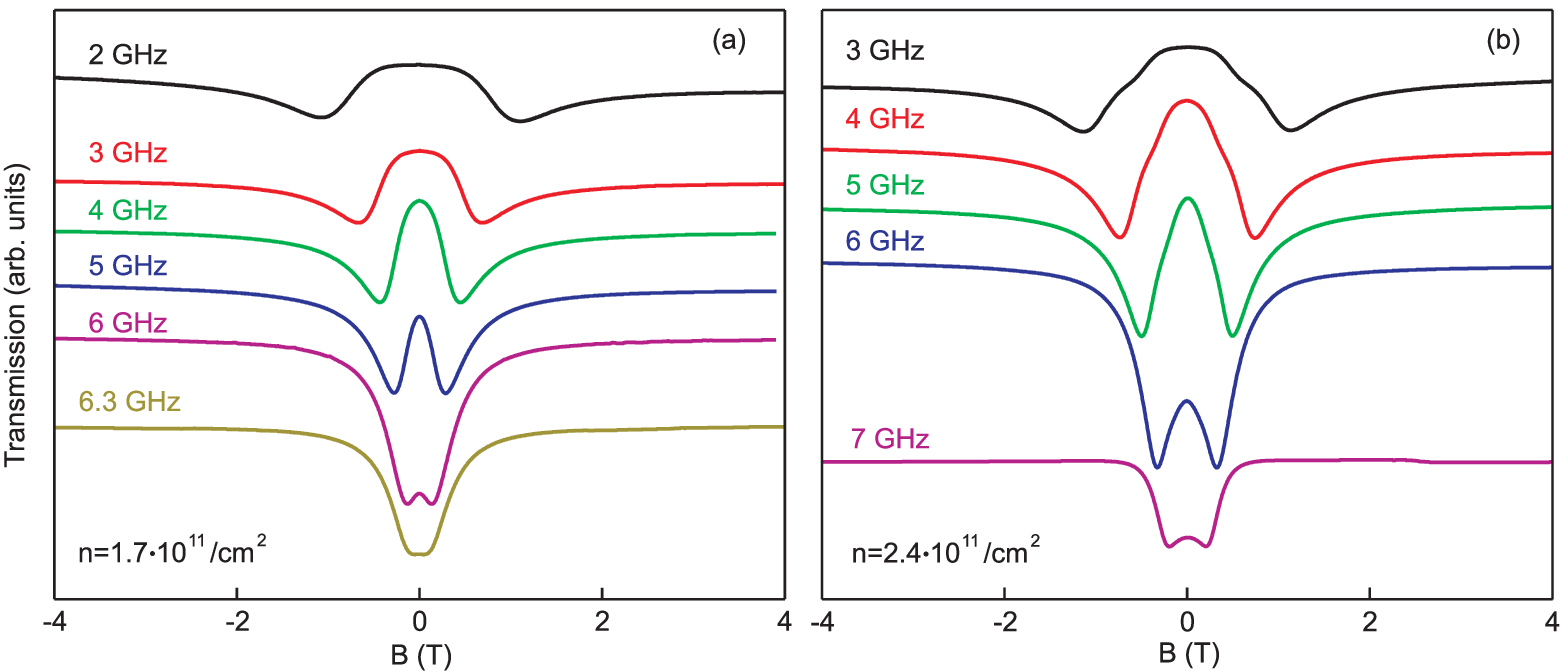}
\caption{Magnetic-field dependencies of the coplanar waveguide transmission for 2DES concentrations (a) $n_s = 1{.}7\times 10^{11}\,\text{cm}^{-2}$ and (b) $n_s = 2{.}4\times 10^{11}\,\text{cm}^{-2}$ at $T=1.5$~K.}
\end{figure}


\section{\textrm{II}. ANALYSIS OF THE RESONANCE LINE-SHAPE}

Part of curves in FIG.~1S(a-b) have an asymmetric line-shape. This asymmetry comes from the fact that transmission through the coplanar waveguide (CPW) is comprised of two interfering contributions $|\vec{E_1}+\vec{E_2}|^2$. One component $\vec{E_1}$ is a broadband transmission of the CPW with disks representing a dissipative load. The second component $\vec{E_2}$ represents the electromagnetic wave that is irradiated by the disks whenever the magnetoplasma resonance occurs. Accordingly, observed asymmetric resonance in the transmission can be described by the Fano resonance line-shape~\cite{Fano}:           
\begin{equation}
\label{fit}
I(B)=C+A \frac{\left(\dfrac{2(B-B_0)}{\Gamma}+q\right)^2}{\left(\dfrac{2(B-B_0)}{\Gamma}\right)^2+1},
\end{equation}
where $B_0$~-- magnetic-field position of the resonance, $\Gamma$ and $A$~-- resonance line-width and amplitude, $C$~-- background level, and $q$~-- asymmetry Fano parameter. 

\begin{figure}[!h]
\includegraphics[width=0.94 \textwidth]{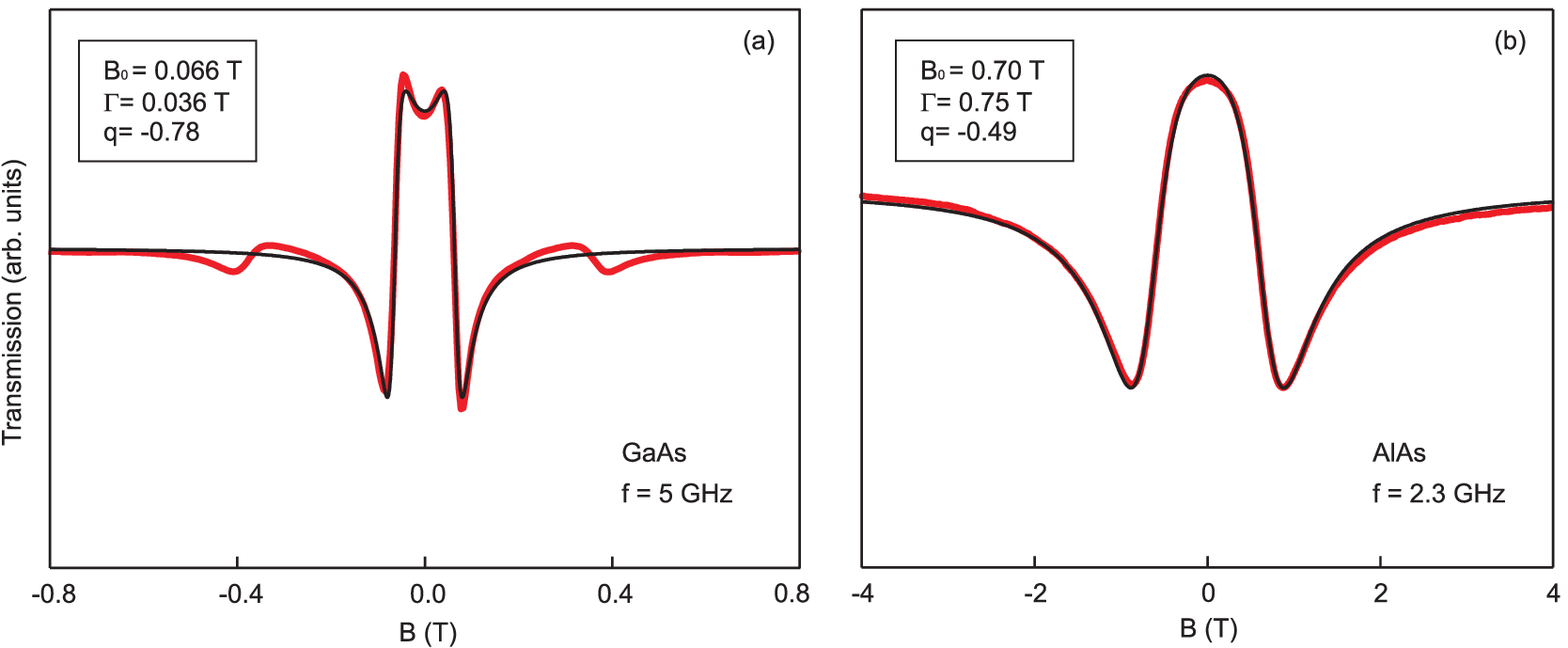}
\caption{Fano resonance fitting for two transmission curves. One for the high-quality GaAs sample with $n_s = 1{.}4\times 10^{11}\,\text{cm}^{-2}$ (a), another for the AlAs sample with $n_s = 1{.}7\times 10^{11}\,\text{cm}^{-2}$ (b).}
\end{figure}

Figure~2S shows the Fano resonance fitting for two transmission curves. One for the high-quality GaAs sample (FIG.~2S(a)), another for the AlAs sample with $n_s = 1{.}7\times 10^{11}\,\text{cm}^{-2}$ (FIG.~2S(b)).